\documentclass[rnote]{aa} % for the research notes
\usepackage{graphicx}
\usepackage{txfonts}
\begin{document}

\title{On the contribution of active galactic nuclei to reionization}

%   \subtitle{}

\author{R.L.~Grissom\and D.R.~Ballantyne\and J.H.~Wise}

\institute{Center for Relativistic Astrophysics, School of Physics,
  Georgia Institute of Technology, 837 State Street, Atlanta, GA 30332-0430, USA\\
\email{rgrissom6@gatech.edu}}

\date{}
 
\abstract{The electron scattering optical depth constraints on reionization suggest that there may be other sources that contribute to the ionization of hydrogen aside from observable star forming galaxies. Often the calculated value of the electron scattering optical depth, $\tau_{es}$, falls below the measurements derived from observations of the CMB or an assumption about non-observable sources must be made in order to reach agreement. Here, we calculate the hydrogen ionization fraction as a function of redshift and the electron scattering optical depth from both galaxies and active galactic nuclei (AGN) factoring in the secondary collisional ionizations from the AGN X-ray emission.  In this paper we use the most current determination of the evolving hard X-ray luminosity function and extrapolate its evolution beyond $z = 6$. The AGN spectral energy distributions (SEDs) include both UV and X-ray ionizing photons. To search for the largest possible effect, all AGN are assumed to have $\lambda_{Edd} = 1.0$ and be completely unobscured. The results show that AGNs produce a perturbative effect on the reionization of hydrogen and remains in agreement with current constraints. Our calculations find the epoch of reionization still ends at $z \approx 6$ and only increases the electron scattering optical depth by  $\sim 2.3\%$ under the most optimal conditions. This can only be moderately increased by assuming a constant black hole mass of $M_{BH} = 10^{5}M_{\odot}$. As a result, we conclude that there is a need for other sources beyond observable galaxies and AGNs that contribute to the reionization of hydrogen at $z > 6$. }


\maketitle
%
%________________________________________________________________

\section{Introduction}
\label{sect:intro}
Most recent studies agree that star forming galaxies are the dominant contributors to the reionization of hydrogen in the universe \citep{aft12, hm12, robertson13}, with other sources, such as active galactic nuclei (AGN) and X-ray binaries, being sub-dominant to this process \citep{willott10, font12, mcquinn12, HB12}. While stars and detectable galaxies satisfy most constraints on the reionization of hydrogen, including ending at $z \approx 6$, there is consistently a disagreement between the calculated and the observed electron scattering optical depth, $\tau_{es}$, with the calculated values falling in the low region of the 68\% range of the measured value. \citet{aft12} calculate a value of $\tau_{es} = 0.086$ (in agreement with the 7-year \textit{WMAP} best fit value; \citealt{wmap7}) under the assumption of two galaxy types: low-mass galaxies with a high escape fraction ($f_{esc} = 0.8$) and high-mass galaxies with low escape fraction ($f_{esc} = 0.05$). The high-mass galaxies dominate reionization for $z > 8$, but are faint and below the current detection limit, while the low-mass galaxies dominate for $z < 8$ \citep{aft12}. When using a constant escape fraction $f_{esc} = 0.2$ for the low-mass galaxies \citet{aft12} find only $\tau_{es} = 0.06$. Similarly,  \citet{ahn12} find $\tau_{es} = 0.0603$, which is well below the 7-year \textit{WMAP} measurements, but by including Pop. III stars as a source, they are able to increase $\tau_{es}$ to 0.0861. \citet{robertson13} determined the reionization history using star-forming galaxies to a limiting magnitude, $M_{UV} < -13$ with an escape fraction of $0.2$ and find $\tau_{es} \approx 0.071$, which is below the best fit value of $0.084 \pm 0.013$ for the electron scattering optical depth from the 9-year \textit{WMAP} measurement \citep{wmap9}. Using a population of star-forming galaxies to a limiting magnitude of $M_{UV} < -17$, which includes only detectable galaxies, yielded an even lower value for $\tau_{es}$. They also considered a population down to a limiting magnitude of $M_{UV} < -10$, which includes a number of undetectable galaxies. This resulted in a value of $\tau_{es}$ closer to the best fit value, but also resulted in reionization ending closer to $z \sim 7$,  further suggesting that there are other sources of reionization beyond low luminosity galaxies.

It is expected that including AGN as a source for reionization will be a perturbative effect rather than a  significant one, but should still increase the calculated electron scattering optical depth. In the \citet{hm12} model of reionization, they conclude that the UV emission from galaxies is the main contributor to reionization of hydrogen sufficient for meeting all existing constraints on this process, including $\tau_{es}$. 
\citet{willott10} calculated that at $z = 6$ the quasar population contributes $\leq$ 20\% of the photons required to maintain ionization. \citet{mcquinn12} concluded that faint quasars could not be solely responsible for the reionization of both hydrogen and helium and remain in agreement with constraints on both histories, specifically the redshifts at which each epoch ends. Alternatively, \citet{volonteri09} modeled the growth and evolution of large and small massive black hole seeds over time and find that at $z \approx 6$ quasars contribute $\leq $ 20\% to the total number of ionizations, but at $z \approx 8$ they could contribute 50\% to 90\% of ionizations.

 Here, we present a new, observationally-motivated calculation to investigate the optimal contribution of AGN as a means of reducing the variation in the calculated and measured values of $\tau_{es}$. The most current parametization of the evolving hard X-ray AGN luminosity function \citep{hiroi12} is used to compute the density of AGN ionizing photons. We also include the effects of secondary collisional ionizations from X-rays as well as the UV emission from star forming galaxies.  Since the focus of this calculation is to determine the maximal contribution to $\tau_{es}$ from AGN, we assume that all AGN accrete at the same constant Eddington rate and are 100\% unobscured. Variations in the Eddington ratio and the X-ray spectral slope are considered to determine the optimal contribution of AGN to the reionization of hydrogen. This simple calculation is independent of models of black hole growth and will provide a new constraint on the AGN contribution to $\tau_{es}$. The cosmological parameters adopted throughout this paper are: $H_0 = 67.77\:\mathrm{km\:s^{-1}\:Mpc^{-1}}, \Omega_m = 0.304, \Omega_{\Lambda} = 0.691, \Omega_b = 0.0483$ \citep{planck13a, planck13b}.  We reproduce the results from \citet{robertson13} for these updated parameters as the baseline for a pure star-forming galaxy source of reionization.

\section{Calculations}
\label{sect:calc}

\subsection{Overview}
\label{sec:over}
The evolution of the comoving ionized fraction of hydrogen, $x(z)$, over time is described by
\begin{equation}
\dot{x} = \frac{\dot{n}(z)}{\langle n_{H}\rangle}-\frac{x}{\langle t_{rec}\rangle},
\label{eq:reionization}
\end{equation}
where $\langle n_H\rangle$, in cm$^{-3}$, is the average, comoving number density of hydrogen atoms, $\dot{n}$ is the number density production rate of total ionizations, and $\langle t_{rec}\rangle$, in s, is the average recombination time. For case B recombinations,
\begin{equation}
\langle t_{rec}\rangle^{-1} = [C_{HII}\alpha_B(T)(1+Y_p/4X_p)\langle n_H\rangle(1+z)^3].
\label{eq:trec}
\end{equation}
$C_{HII}$ is the clumping factor quantifying the inhomogeneity in the intergalactic medium (IGM), $\alpha_B(T)$ is the case B recombination coefficient, and $X_P$ and $Y_P$ are the present hydrogen and helium abundances, respectively. Following \citet{robertson13}, the IGM temperature is assumed to be 20,000 K, giving $\alpha_B = 2.59 \times 10^{-13}\:\mathrm{cm^3 \: s^{-1}}$ and $C_{HII} =3.0$ \citep{pawlik09}.

The time-dependent production rate of ionizations,  $\dot{n}(z)$, depends on the source of ionizing photons. When considering an ionizing source of both star forming galaxies and AGN, the ionization production rate is $\dot{n}=\dot{n}_{Gal}+\dot{n}_{AGN}$. For star forming galaxies, $\dot{n}_{Gal}$ can be estimated as, 
\begin{equation}
\dot{n}_{Gal} = f_{esc}\xi_{ion}\rho_{UV}
\label{eq:nionGAL}
\end{equation}
where $f_{esc}$ is the escape fraction of ionizing photons, $\xi_{ion}$ is the hydrogen ionizing photons per 1500$\mathring{\mathrm{A}}$ luminosity, and $\rho_{UV}$ is the integrated luminosity density down to a limiting magnitude, $M_{UV} < -13$ (note that this does include objects below the current detection limits). From \citet{robertson13}, $f_{esc} = 0.2$ and $\xi_{ion} = 10^{25.2}\:\mathrm{Hz \: erg^{-1}}$.

\subsection{AGN Ionization Production Rate}
\label{sec:nionAGN}
The  AGN contribution is defined
\begin{equation}
\dot{n}_{AGN} = \int_{\log L_{X} \: \mathrm{(min)}}^{\log L_{X} \: \mathrm{(max)}} N(x, L_{X}) \frac{d\Phi \left(z, \log L_X\right)}{d\log L_{X}} d\log L_{X},
\label{eq:nionAGN}
\end{equation}
with $N(x, L_{X})$, in $\mathrm{s^{-1}}$, giving the total ionizations (including secondary collisional ionizations) occurring per second for a single AGN with rest-frame X-ray luminosity $L_{X} = L_{2-10\mathrm{keV}}$ (in $\mathrm{erg \: s^{-1}}$) and surrounded by a medium in which the fraction of ionized hydrogen is $x$. The limits of the integration are $\log L_X \: \mathrm{(min)} = 41.5$ and $\log L_X \: \mathrm{(max)} = 48$. We adopt the X-ray luminosity function $\left(d\Phi/d\log L_{X}\right)$ from the work of \citet{hiroi12} who determined the power law decline at higher redshifts. Note  that while this model is derived based on observable data to redshift $z \approx 4.5$ \citep{hiroi12} there are no data to confirm the accuracy of this luminosity function extrapolated out to redshift $z = 30$. However, below we compare our $\dot{n}_{AGN}$ at $z = 6$ with \citet{willott10} to justify this extrapolation.

To calculate the ionization rate produced by a single AGN,
\begin{equation}
N(x, L_X) = \int_{13.6\mathrm{eV}}^{1000\mathrm{keV}} \frac{L_E}{E}(1+\chi(x, E)) dE,
\label{eq:Ninztns}
\end{equation}
where $L_E(E)$ is the spectral energy distribution with units $\mathrm{erg \: s^{-1} \: keV^{-1}}$. Since the energy spectrum of an AGN includes a significant number of high energy photons, the total number of ionizations per photon with initial energy, $E$, and background ionized hydrogen fraction, $x$, depends on the number of secondary collisional ionizations, $\chi(x, E)$. \citet{schull85} found that for a photon with initial energy $E \geq 100$ eV in a medium with an ionized hydrogen fraction, $x$, the number of secondary collisional ionizations of hydrogen is best described as
\begin{equation}
\centering
\chi(x, E) = 1 + C(1-x^{a})^{b}  \frac{E}{E_I}.
\label{eq:totalioniz}
\end{equation}
with parameters $C = 0.3908, a = 0.4092,$ and $b = 1.7592$ and an ionization energy $E_I = 13.6$ eV. This analytic fit is only valid for an initial energy $\geq 100$ eV.  As the simulations also show that $\chi$ is a linear function of energy, $E$, for a fixed ionized fraction, $x$ \citep{schull85}, we linearly extrapolate this relationship for energies $E \leq 100$ eV.

\begin{figure}
\centering
\includegraphics[width=0.5\textwidth]{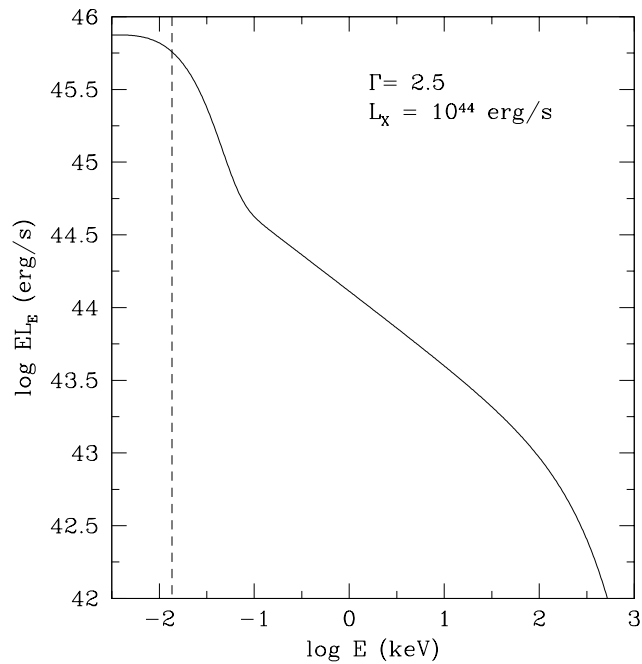}
\caption{General form of the spectral energy distribution for a single AGN given by Equation (\ref{eq:SED}) with an accretion rate of $\lambda_{Edd} = 1.0$.  For this particular AGN, $\Gamma = 2.5$ and the UV to X-ray correlation parameter is $\alpha_{ox} = -1.73$. The dashed line represents $E = 13.6$ eV.}
\label{fig:SED}
\end{figure}

To construct the SED for an AGN with a given X-ray luminosity $L_X$, we start with the following general shape:
\begin{equation}
L_E(E) = N \left[ \left(\frac{E}{h}\right)^{\alpha_{UV}} \exp{(-E / k T_{BB})} + a \left(\frac{E}{h}\right)^{1-\Gamma} \exp{(-E/E_{cut})}\right].
\label{eq:SED}
\end{equation}
The first term represents the thermal accretion disk spectrum, where $\alpha_{UV} = -0.5$ is the UV spectral slope and $T_{BB}$ is the characteristic blackbody temperature of the disk. The second term is the X-ray power law spectrum, where $\Gamma$ is the X-ray photon index and $N$ is an overall normalization factor in order to give the needed $L_X$. The exponential in the second term is a high energy cut-off $E_{cut} = 300$ keV e.g.,\citep[e.g, ][]{ricci11}.  

To connect the X-ray power law to the optical emission, we use the $\alpha_{ox}$ parameter,
\begin{equation}
\alpha_{ox} = \frac{\log\left(L_{2\mathrm{kev}} / L_{2500\mathring{\mathrm{A}}}\right)}{ \log\left(\nu_{2\mathrm{keV}} / \nu_{2500\mathring{\mathrm{A}}}\right)},
\label{eq:aox}
\end{equation}
which defines the slope between the optical spectrum and the X-ray spectrum. As $\alpha_{ox}$ is a measure of the relative strength of the X-ray emission it is observed to be correlated with $\kappa_{Bol}=L_{Bol}/L_{X}$, the X-ray bolometric correction \citep{lusso10}. $\kappa_{Bol}$, in turn, is related to $\lambda_{Edd} = L_{Bol}/L_{Edd}$ \citep{lusso10}.  $L_{Bol}$ and $L_{Edd}$ are the bolometric luminosity and the Eddington luminosity, respectively. Thus, for a given $L_X$ and $\lambda_{Edd}$ a corresponding $\kappa_{Bol}$ and $\alpha_{ox}$ are calculated. From the bolometric correction, the black hole mass is determined via $L_{Edd} =(1.38\times10^{38} )(M_{BH}/M_{\odot})\mathrm{erg \: s^{-1}}$. This yields the characteristic blackbody temperature, $T_{BB} = 1.37\times10^{7}\left(M_{BH}/M_{\odot}\right)^{-1/4}$ K, needed to determine the optical bump in the SED. Fig. \ref{fig:SED} shows an example SED. As a check, this SED is integrated from $1\:\mu$m to 1000 keV to calculate a bolometric luminosity of $10^{46.4}\:\mathrm{erg\:s^{-1}}$m, which is only 2.5 times larger than the empirical luminosity, $L_{Bol} = \kappa_{Bol}L_{X} = 10^{46}\:\mathrm{erg\:s^{-1}}$. Given the uncertainties in measuring $\kappa_{Bol}$ and the true AGN SED, this result supports the methods used in this paper.

The SED in our calculation is defined entirely by three parameters: $L_X$, $\Gamma$, and $\lambda_{Edd}$. Given values for these quantities, we can construct an SED for any AGN. It should be noted that for the entire population the ideal assumption is made that all AGN accrete at the same $\lambda_{Edd}$. 

\begin{figure}[t!]
\centering
\includegraphics[width=0.5\textwidth]{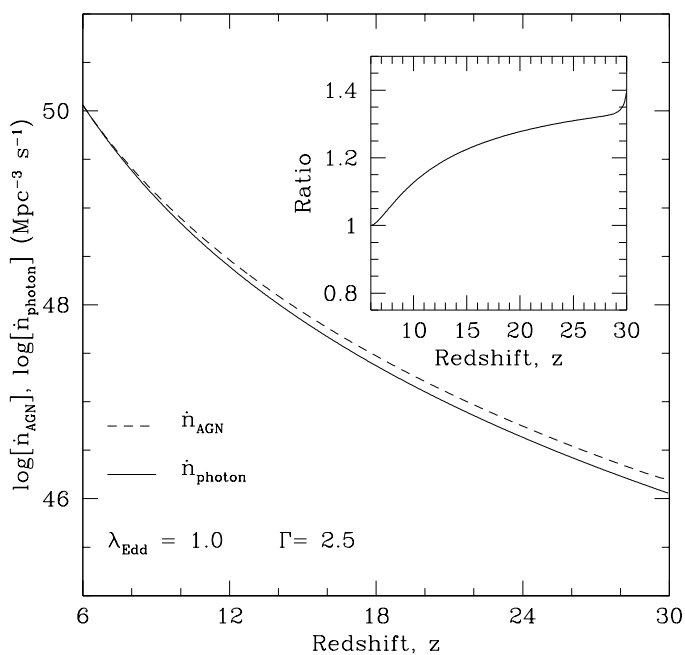}
\caption{Comparison of the production rate of ionizing photons to the rate of actual ionizations, including secondary collisional ionizations. This is the maximum ideal case where all AGN are assumed to accrete at a rate of $\lambda_{Edd} = 1.0$ and the X-ray photon index is $\Gamma = 2.5$. The difference between the two curves increases as redshift increases, where the IGM becomes more and more neutral. The hydrogen ionization fraction, $x$, used to calculate $\dot{n}_{AGN}$ by Equation \ref{eq:nionAGN} is that due to star forming galaxies calculated by \citet{robertson13}. Important to note that even though the high energy photons cause additional ionizations, the number remains below $\dot{n} \approx 2 - 4 \times 10^{50}\mathrm{Mpc^{-3}}$ as seen in the work of \citet{aft12} for the galactic ionization production rate. The inset shows the ratio of ionizations to ionizing photons.}
\label{fig:ninztnsAGN}
\end{figure}

Once the SED is computed from Equation (\ref{eq:SED}), then $\dot{n}_{AGN}$ can be calculated by Equations (\ref{eq:Ninztns}) and (\ref{eq:totalioniz}). As an example, Fig. \ref{fig:ninztnsAGN} shows the ionizing photon production rate solely from AGN, $\dot{n}_{photon}$, versus the AGN ionization production rate, $\dot{n}_{AGN}$, for an accretion rate of $\lambda_{Edd} = 1.0$ and $\Gamma = 2.5$. Though we consider various values of $\Gamma$ for a fixed $\lambda_{Edd} = 1.0$, \citet{bright13} determined there is a strong correlation between $\Gamma$ and $\lambda_{Edd}$; for $\lambda_{Edd} = 1.0$, $\Gamma \approx 2.3$. As a result, we consider $\lambda_{Edd} = 1.0$ and $\Gamma = 2.5$ to be our ideal case. Since the number of total ionizations is directly dependent on the ionized fraction at any given redshift, $z$, the evolution of $x$ due to star forming galaxies from \citet{robertson13} is used to produce $\dot{n}_{AGN}$. Although this is not entirely self-consistent, it is known that AGN do not change the evolution of $x(z)$ significantly, so it is sufficient to use the \citet{robertson13} results to demonstrate the effects of secondary collisional ionizations. As expected, at higher redshift when the ionization fraction drops, there appears to be more ionizations than photons as a result of collisional ionizations within a partially-ionized medium. By decreasing $\lambda_{Edd}$ from 1.0 to 0.1 the ratio of ionizations to ionizing photons at $z = 10$ increases from $\approx 1.1$ to $\approx 1.25$. This increase in collisional ionizations results from the higher contribution of X-rays to the bolometric luminosity based on the definitions of $\lambda_{Edd}$ and $\kappa_{bol}$. However,  though there maybe 15\% more ionizations compared to the number of available ionizing photons for a lower Eddington ratio, the total number of ionizing photon decreases overall. Changing the photon index, on the other hand, has very little effect. The minimum ratio occurs for $\Gamma = 2.0$ and increases only slightly by either decreasing or increasing $\Gamma$.

For this highly idealized model of an AGN source, $\log \dot{n}_{AGN} \approx 50.1$ at $z = 6$ for the optimal scenario of $\Gamma = 2.5$ and $\lambda_{Edd} = 1.0$. Checking this against the work of \citet{willott10}, in which they derived the luminosity function at redshift 6 using a sample of 40 quasars, our results show a slightly higher ionizing photon production rate than their maximum allowed value of $\log \dot{n}_{AGN} = 49.4$. Yet, our value for $\dot{n}_{AGN}$ is still within a factor of 5 of their upper bound estimate. Given the simplifications made in our model, this is very reasonable.

Equation (\ref{eq:nionAGN}) and Equation (\ref{eq:reionization}) are solved simultaneously using a fourth-order Runge-Kutta method to calculate the fraction of ionized hydrogen. At redshift $z = 30$, $x$ is assumed to be $10^{-5}$ for the initial conditions. Though an escape fraction is assumed for the photon emissivity from the galaxies, no such constraints are made for the AGN. Assuming 100\% of ionizing photons escape from all AGN means that obscuration is not being taken into account and that the results given in the next section are idealized maximum cases.

\subsection{Electron Scattering Optical Depth}
\label{sec:optDepth}
One of the main constraints on the reionization history is the electron scattering optical depth of the IGM. Using the ionization fraction, $x(z)$, found by solving Equation (\ref{eq:reionization}) the optical depth is computed as
\begin{equation}
\tau_{es} = \int_0^{30} x(z) \frac{c \sigma_T \langle n_H \rangle f_e (1+z)^2}{H_0 \sqrt{\Omega_m(1+z)^3+\Omega_{\Lambda}}}dz
\label{eq:optDepth}
\end{equation}
where $c$ is the speed of light, $\sigma_T$ is the Thompson cross section, and $f_e$ is the number of free electrons in the IGM per a hydrogen nucleus. The last quantity depends on the ionization of both helium and hydrogen. \citet{aft12} defined the break redshift at which helium goes from being singly ionized to doubly ionized at $z = 3$ and \citet{robertson13} define this point to be at $z = 4$. In keeping consistent with the work of \citet{robertson13}, $f_e = (1+Y_p/2 X_p)$ for $z \leq 4$ and $f_e = (1+Y_p/4 X_p)$ for $z > 4$.

\section{Results}
\label{sect:res}
\subsection{Contribution of AGN to the Ionized Hydrogen Fraction}
\label{sec:AGNreionize}
% Two figures across both columns
\begin{figure*}
\centering
\includegraphics[width=0.4\textwidth]{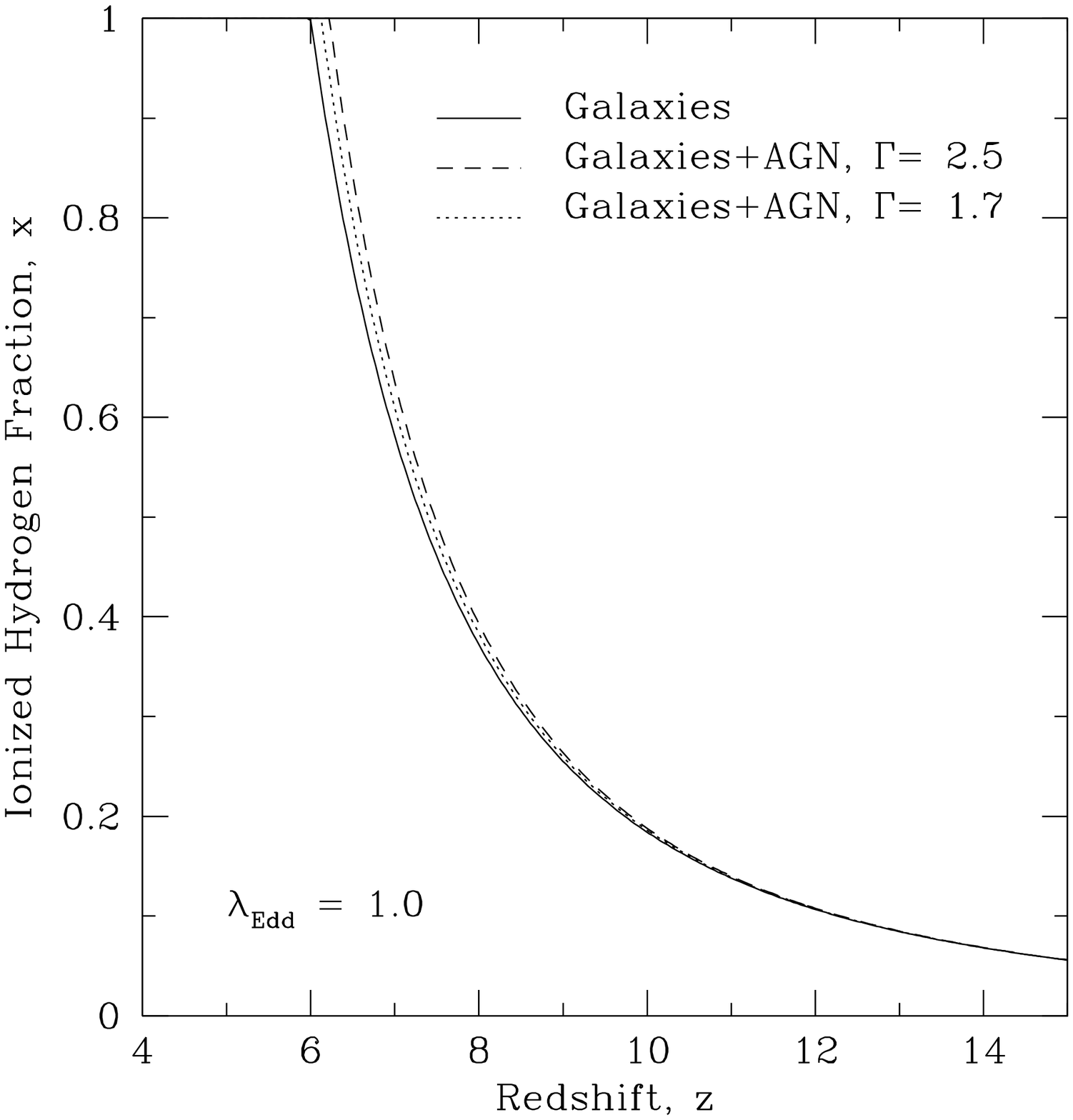}
\includegraphics[width=0.4\textwidth]{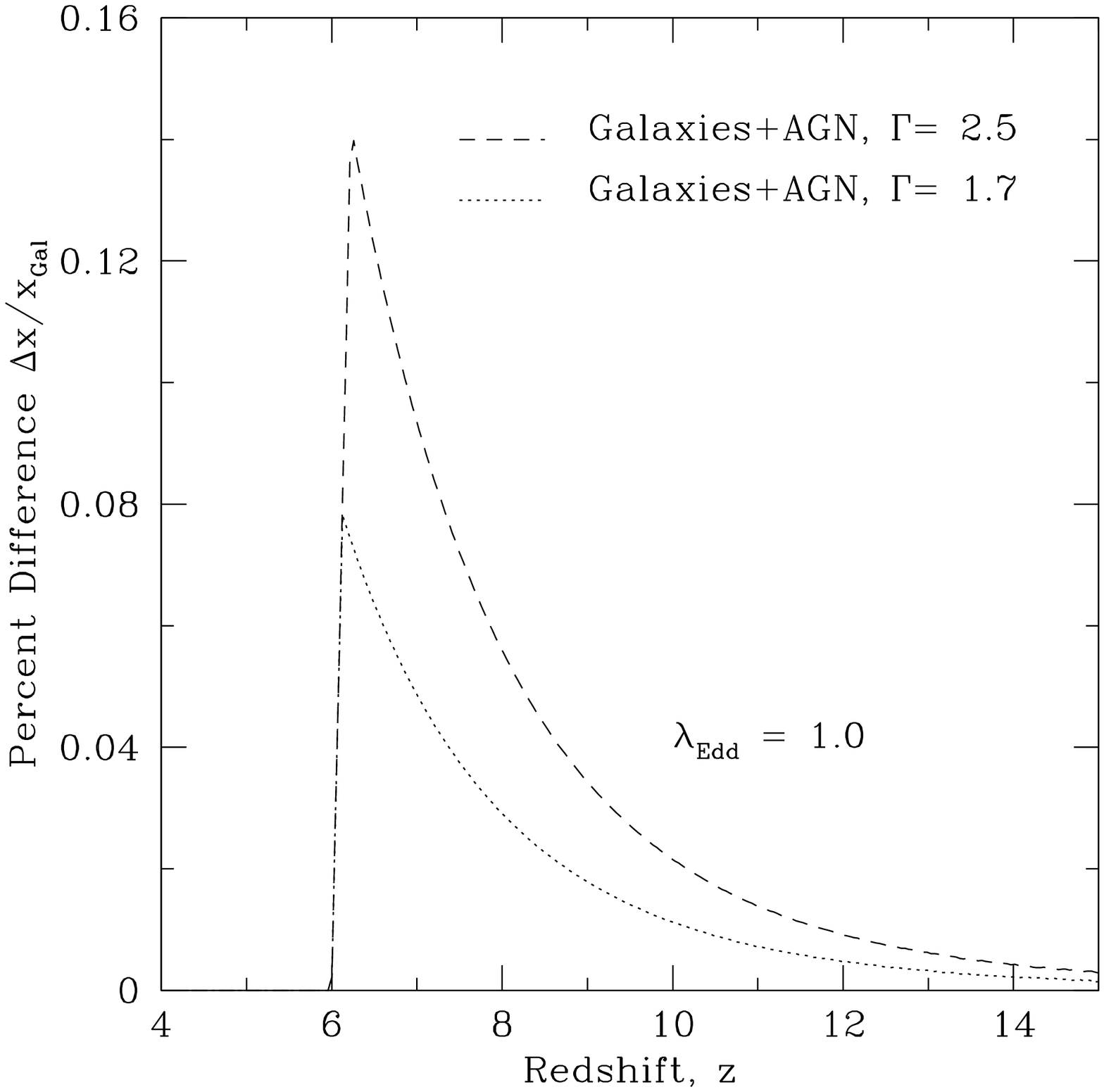}
\caption{Left panel gives the ionized hydrogen fraction $x(z)$ for (1) a star forming galaxy source \citep{robertson13} (solid line), (2) galaxies plus AGN with $\lambda_{Edd} = 1.0$ and $\Gamma = 2.5$ (long dashed line), and (3) galaxies plus AGN with $\lambda_{Edd} = 1.0$ and $\Gamma = 1.7$ (dotted line). All three cases show reionization ending around $z \sim 6$. The right panel shows the percent difference in the ionization fraction between each AGN case and the galaxy case as a function of redshift. In the optimal case, there is a maximum of 14\% increase in $x$ near the end of reionization. }
\label{fig:reionization}
\end{figure*}

The left panel in Fig. \ref{fig:reionization} shows the ionization fraction found by solving Equation (\ref{eq:reionization}) for three different sources of ionization: the \citet{robertson13} results for galaxies, galaxies and AGN with $\lambda_{Edd} = 1.0$  and $\Gamma = 2.5$, and galaxies and AGN with $\lambda_{Edd} = 1.0$ and $\Gamma = 1.7$.  As expected from the previous works of \citet{font12}, \citet{willott10}, and \citet{hm12}, the addition of AGN appears to be a perturbation of the galactic contribution, with no dramatic change in the time at which reionization ends. Without an AGN contribution to reionization, hydrogen is completely ionized at $z = 6$. In the optimal case, $\lambda_{Edd} = 1.0$ and $\Gamma = 2.5$, reionization ends at $z = 6.26$. It ends at $z = 6.13$ for $\Gamma = 1.7$.

To better observe the effects of AGN on reionization, the percent difference ($\Delta x / x = (x_{AGN} - x_{Gal}) / x_{Gal}$) is calculated and shown in the right panel of Figure (\ref{fig:reionization}). A maximum 14\% increase in the ionization fraction of hydrogen is observed near $z \sim 6$ for $\Gamma = 2.5$ and $\lambda_{Edd} = 1.0$ and only a maximum 8\% increase is observed around the same time for $\Gamma = 1.7$. Decreasing the Eddington ratio to 0.1 reduces the effect of the AGN source from a  $\approx 10\%$ increase to a $\approx 1\%$ increase (for varying values of $\Gamma$) in $x$ near reionization.

\subsection{Electron Scattering Optical Depth}
\label{sec:AGNtau}
\begin{figure}
\centering
\includegraphics[width=0.5\textwidth]{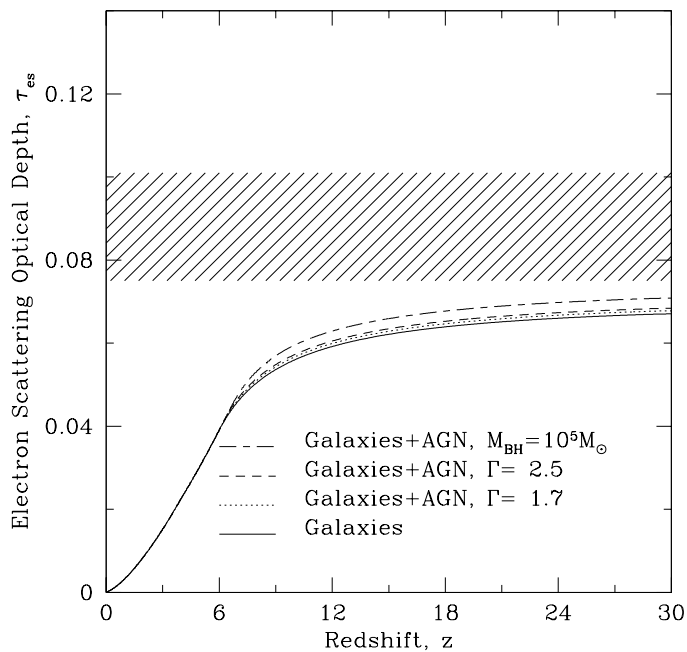}
\caption{Electron scattering optical depth for the same three cases as in Fig. \ref{fig:reionization}: (1) a star forming galaxy source \citep{robertson13} (solid line), (2) galaxies plus AGN with $\lambda_{Edd} = 1.0$ and $\Gamma = 2.5$ (long dashed line), and (3) galaxies plus AGN with $\lambda_{Edd} = 1.0$ and $\Gamma = 1.7$ (dotted line). Also shown is the evolution of $\tau_{es}$ for galaxies plus an AGN population with a constant black hole mass $M_{BH} = 10^5M_{\odot}$. The shaded region represents the most tightly constrained 68\% margins of $\tau_{es}$ as measured by \citet{planck13b}. All four cases fall below this region and, thus, below the best fit value of 0.089.}
\label{fig:opticaldepth}
\end{figure}

The effects of AGN on $\tau_{es}$ are shown in Fig. \ref{fig:opticaldepth}. For star forming galaxies of \citet{robertson13}, $\tau_{es} = 0.067$ when calculated over $0 \leq z \leq 30$. Note this value is slightly lower than $\tau_{es} \approx 0.071$ reported by \citet{robertson13}. This is mainly due to the use of the \textit{Planck} cosmological parameters instead of the 9-year WMAP parameters employed by \citet{robertson13}. Including AGN with an accretion rate of $\lambda_{Edd} = 1.0$ increases $\tau_{es}$ to 0.0684 for $\Gamma = 2.5$ and to 0.0678 for $\Gamma = 1.7$. These results are equivalent to an overall $\sim 2.3\%$ increase and $\sim 1.1\%$ increase, respectively and represent the maximum contributions observed for our idealized model of AGN as an ionizing source. This further supports the observation that an AGN source of reionization acts as a perturbation of the dominant galactic contribution.

Compared to the measurement of $\tau_{es} = 0.089^{+ 0.012}_{-0.014}$ \citep{planck13b} derived from combining the \textit{Planck} observations and \textit{WMAP} polarization, all of the results shown in Fig. \ref{fig:opticaldepth} fall below the best fit value and lower end of this range. This suggests there may be other sources of reionization. \citet{aft12} studied the possibility of high mass, low luminous galaxies that fall below the current detection limit as a main contribution to reionization at high redshifts. Their work reproduces the measured \textit{WMAP + Planck} constraint on $\tau_{es}$. \cite{robertson13} also investigated the contribution of faint galaxies (below the current detectable limit) for three different limiting magnitudes, $M_{UV}$: $-10$, $-13$, and $-17$. They find that considering only the detectable galaxies ($M_{UV} < -17$) can't reionize the Universe by $z \approx 6$ and $\tau_{es} \approx 0.05$, falling well below the \textit{Planck} results. Considering galaxies down to $M_{UV} = -10$ yields $\tau_{es} \approx 0.075$, which just lies within the 68\% margin of the \textit{Planck} measurements, but reionization ends closer to $z = 7$. Using binary population synthesis simulations and observations from present to $z \approx 4$, \citet{frag13} studied the evolution of X-ray binaries (XRBs) and their high energy contributions to the IGM. Their results suggest that while AGN are the dominant X-ray source for $0 \leq z \leq 4$, XRBs dominate the X-ray energy contribution to the IGM for $6 \leq z \leq 8$ and most likely for higher redshifts as well. Thus, it's possible that these XRBs contribute more significantly to reionization than AGN. Simulating reionization with and without the very first stars formed in the Universe, \citet{ahn12} found that these stars contribute $\sim 0.1 -0.2$ to the electron scattering optical depth. Including Pop. III brings their calculated value of $\tau_{es}$ to a value in agreement with \textit{Planck}. The low values of $\tau_{es}$ calculated in this paper for just star forming galaxies with $M_{UV} \leq -13$ and AGN give support to these areas of study.

\subsection{Constant $M_{BH}$}
\label{sec:constBHM}
We also briefly consider the scenario in which all AGN have the same constant black hole mass, $M_{BH} = 10^{5}M_{\odot}$. Though this is not accurate at $z = 6$, for higher redshifts ($z \geq 8$) this may be a more plausible scenario \citep{volonteri09}. An AGN population with a constant black hole mass of $10^{5}M_{\odot}$ has a higher accretion disk temperature and, thus, produces more ionizing photons than a population of AGN with just a constant $\lambda_{Edd}$.

Solving Equation (\ref{eq:reionization}) assuming a constant black hole mass $M_{BH} = 10^{5}M_{\odot}$, an $\alpha_{ox}$ consistent with $\lambda_{Edd} = 1.0$, and $\Gamma = 2.5$ for all AGN, we find that reionization ends at $z = 6.6$, earlier than any of the results discussed in \ref{sec:AGNreionize}. This is evidence that a constant black hole mass is not viable around $z \sim 6$. Under these same conditions there is a 40\% increase in the ionization fraction around $z = 6$, almost 3 times as great as the optimal increase mentioned in Section \ref{sec:AGNreionize}. 

The contribution of AGN under this particular case also increases the electron scattering optical depth to $\tau_{es} = 0.071$ corresponding to a $5.8\%$ increase in $\tau_{es}$ when compared to just the contribution from star forming galaxies as in \citet{robertson13}. Again, this increase is approximately 2.5 times greater than the optimal increase observed in Section \ref{sec:AGNreionize}, but still not significant enough to bring the calculated value of $\tau_{es}$ in agreement with the \textit{Planck} measurements, providing evidence that further sources of reionization are needed. To finally settle the contribution of AGNs requires measures of quasar luminosity functions at $z \geq 6$ by, e.g., LSST.

% Two figures across both columns

\section{Summary}
\label{sec:sum}
We quantitatively evaluate the contribution of AGNs to reionization using the latest evolving hard X-ray AGN luminosity function, the consideration of secondary ionizations, and different values of $\lambda_{Edd}$ and $\Gamma$. The effects of high energy X-ray photons from the AGN are more perturbative than significant, in agreement with previous works \citep{willott10, font12, hm12, mcquinn12}. Here, we consider an optimal, idealized model of AGN reionization; i.e. $\lambda_{Edd} = 1.0$ for all AGN, no obscuration. Under these maximal conditions our results show reionization still ends at $z \approx 6$, but $\tau_{es}$ only increases by $\leq 2.3\%$, below the current \textit{Planck} measurement. Despite looking at changes in $\Gamma$ and $\lambda_{Edd}$, we find the only significant change in $\tau_{es}$ (an increase of $\approx 5.8\%$) follows from assuming a constant black hole mass of $10^5M_{\odot}$ for the entire AGN population. However, this still yields a $\tau_{es}$ below the \textit{Planck} value of $0.089^{+ 0.012}_{-0.014}$ \citep{planck13b}. It may be possible for AGN to have a greater contribution to the reionization of hydrogen if the evolution of the AGN X-ray luminosity function dramatically alters at $z \sim 10$, but there is currently not enough observational data to support this type of evolution. Based on the small increase of $\tau_{es}$, even with the optimal assumption of parameters to constrain this calculation, there is a need for further perturbative sources to reionization at $z > 6$ other than AGN, such as X-ray binaries, Pop. III stars, and/or star-forming galaxies that lie below current detection limits.

\begin{acknowledgements}
This work was supported in part by NSF award AST 1008067 to DRB and AST 1211626 to JHW.
\end{acknowledgements}

%-------------------------------------------------------------------


\begin{thebibliography}{}
\bibitem[\protect\citeauthoryear{Ade et al.}{2013a}]{planck13a}
  Ade, P.A.R., Aghanim, N., Armitage-Caplan, C., et al., 2013, \aap, in press (arXiv: 1303.5062)
\bibitem[\protect\citeauthoryear{Ade et al.}{2013b}]{planck13b}
  Ade, P.A.R., Aghanim, N., Armitage-Caplan, C., et al., 2013, \aap, in press (arXiv: 1303.5076)
\bibitem[\protect\citeauthoryear{Ahn et al.}{2012}]{ahn12}
  Ahn, K., Iliev, I.T., Shapiro, R., et al., 2012, \apj, 756, L16
\bibitem[\protect\citeauthoryear{Alvarez et al.}{2012}]{aft12}
  Alvarez, M.A., Finlator, K. \& Trenti, M., 2012, \apj, 759, L38
\bibitem[\protect\citeauthoryear{Brightman et al.}{2013}]{bright13}
  Brightman, M., Silverman, J.D., Mainieri, V., et al., 2013, \mnras, 433, 2485
\bibitem[\protect\citeauthoryear{Dunkley et al.}{2009}]{dunk09}
  Dunkley, J., Komatsu, E., Nolta, M.R., et al., 2009, \apjs, 180, 306
\bibitem[\protect\citeauthoryear{Fontanot et al.}{2012}]{font12}
  Fontanot, F., Cristiani, S., \& Vanzella, E., 2012, \mnras, 425, 1413
\bibitem[\protect\citeauthoryear{Fragos et al.}{2013}]{frag13}
  Fragros, T., Lehmer, B.D., Naoz, S., et al., 2013, \apj, 776, L31
\bibitem[\protect\citeauthoryear{Haardt \& Madau}{2012}]{hm12}
  Haardt, F. \& Madau, P., 2012, \apj, 746, 125
\bibitem[\protect\citeauthoryear{Hinshaw et al.}{2012}]{wmap9}
  Hinshaw, G., Larson, D., Komatsu, E., et al., 2012, \apjs, 208, 19
\bibitem[\protect\citeauthoryear{Hiroi et al.}{2012}]{hiroi12}
  Hiroi, K., Ueda, Y., Akiyama, M., \& Watson, M., 2012, \apj, 758, 49
\bibitem[\protect\citeauthoryear{Holley-Bockelmann et al.}{2012}]{HB12}
  HolleyBockelmann, K., Wise, J.H., \& Sinha, M., 2012, \apj, 761, L8
\bibitem[\protect\citeauthoryear{Komatsu et al.}{2011}]{wmap7}
  Komatsu, E., Smith, K., Dunkley, J., et al., 2011, \apjs, 192, 18
\bibitem[\protect\citeauthoryear{Lusso et al.}{2010}]{lusso10}
  Lusso, E., Comastri, A., Vignali, C., et al., 2010, \aap, 512, A34
\bibitem[\protect\citeauthoryear{McQuinn}{2012}]{mcquinn12}
  McQuinn, M., 2012, \mnras, 426, 1349
\bibitem[\protect\citeauthoryear{Pawlik et al.}{2009}]{pawlik09}
  Pawlik, A.H., Schaye, J., \&, van Scherpenzeel, E., 2009, \mnras, 394,1812
\bibitem[\protect\citeauthoryear{Ricci et al.}{2011}]{ricci11}
  Ricci, C., Walter, R., Courvoisier, T., et al., 2011, \aap, 532, A102
\bibitem[\protect\citeauthoryear{Robertson et al.}{2013}]{robertson13}
  Robertson, B.E., Furlanetto, S.R., Schneider, E., et al., 2013,
  \apj, 768, 71
\bibitem[\protect\citeauthoryear{Shull \& van Steenberg}{1985}]{schull85}
  Shull, J.M., \& van Steenberg, M.E., 1985, \apj, 298, 268
\bibitem[\protect\citeauthoryear{Volonteri \& Gnedin}{2009}]{volonteri09}
  Volonteri, M. \& Gnedin, N.Y., 2009, \apj, 703, 2113
\bibitem[\protect\citeauthoryear{Willott et al.}{2010}]{willott10}
  Willott, C.J., Delorme, P., Reyl\'{e}, C., et al., 2010, \aj, 139, 906
  

\end{thebibliography}
\end{document}